\documentstyle[12pt,epsfig]{article}

\newcommand{\mrm}[1]{\mbox{\rm #1}}

\newcommand{\be}{\begin{equation}}
\newcommand{\ee}{\end{equation}}

\newcommand{\bea}{\begin{eqnarray}}
\newcommand{\eea}{\end{eqnarray}}

\newcommand{\gsim}{\ \rlap{\raise 2pt\hbox{$>$}}{\lower 2pt \hbox{$\sim$}}\ }
\newcommand{\lsim}{\ \rlap{\raise 2pt\hbox{$<$}}{\lower 2pt \hbox{$\sim$}}\ }

\newcommand{\ssu}{$SU(2)_L\times SU(2)_R\times U(1)_{B-L}\,$}

\newcommand{\sulu}{$SU(2)_L\times U(1)_Y$}

\newcommand{\matr}{\left( \begin{array}}
\newcommand{\ematr}{\end{array} \right)}
\newcommand{\D}{\Delta}
\newcommand{\g}{\gamma}

\makeatletter
\if@twoside
\else \oddsidemargin 00pt \evensidemargin 00pt
\fi
\let\@eqnsel = \hfil

\expandafter\ifx\csname mathrm\endcsname\relax\def\mathrm#1{{\rm #1}}\fi
\@ifundefined{mathrm}{\def\mathrm#1{{\rm #1}}}{\relax}

\makeatother

\begin{document}
\thispagestyle{empty}
\null
\hfill FTUV/96-63, IFIC/96-71

\hfill HU-SEFT R 1996-24

\hfill hep-ph/9611362

\vskip 1.5cm

\begin{center}
{\Large \bf      
CONSTRAINTS ON DOUBLY CHARGED HIGGS INTERACTIONS AT LINEAR COLLIDER
\par} \vskip 2.em
{\large		
{\sc G. Barenboim$^{1}$, K. Huitu$^2$ J. Maalampi $^3$ and M. Raidal$^{1}$
}  \\[1ex] 
{\it $^1$ Departament de F\'\i sica Te\`orica, Universitat 
de Val\`encia 
\\  E-46100 Burjassot, Valencia, Spain} \\[1ex]
{\it $^2$ Research Institute for High Energy Physics, \\  
FIN-00014 University of Helsinki, Finland} \\[1ex]
{\it $^3$ Theory Division, Department of Physics, \\ 
FIN-00014 University of Helsinki, Finland}\\[1ex]

\vskip 0.5em
\par} 
\end{center} \par
\vfil
{\bf Abstract} \par
Production of a single doubly charged Higgs boson $\D^{--}$
in polarized $e^+e^-$ and $e^+\g$ collision modes of the linear collider 
have been investigated. The mass range of $\D^{--}$ to be probed 
extends up to the collision energy. 
The diagonal lepton number violating Yukawa
coupling $h_{ee}$ will be tested at least three orders of magnitude 
more strictly than in present experiments.
\par
\vskip 0.5cm
\noindent October 1996 \par
\null
\setcounter{page}{0}
\clearpage

1. {\it Introduction.} Discovery of a
doubly charged scalar particle in  future colliders 
would be a definite signal of  new physics beyond 
the Standard Model (SM).
One of the most attractive theories in which such particles are present is
the left-right symmetric (LR)  electroweak theory \cite{pati}. 
This theory, based on the gauge symmetry group \ssu, was proposed to offer a
dynamical solution to the parity violation  of weak interaction. 
The presence of  triplet representations of Higgs fields, i.e. 
 SU(2)$_R$ triplet $\Delta_R$ and  SU(2)$_L$ triplet $\Delta_L$,
 provides a simple explanation to  
the lightness of   neutrinos via the see-saw mechanism \cite{seesaw}.
The triplet scalars do not couple to quarks and their couplings to leptons 
break
the lepton number by two  units, leading 
to a clear decay signature of the doubly charged scalars, namely 
a same sign pair of leptons. 

In literature several experimental tests of 
lepton number violating interactions
mediated by  virtual doubly charged 
bosons have been reported. There are  two unknown
parameters on which the obtained constraints depend: the mass of the scalar
$M_{\Delta^{--}}$ and a coupling constant
$h_{ij}$, where $i,j=e, \mu$ 
(no constraints are available for the $\tau$ leptons).
Assuming that the rest energy of 
the scalar is large compared with the interaction
energy, the constraints one can derive 
from the present measurements are upper limits
for  quantities of the type $h_{ij}h_{i'j'} /M_{\Delta^{--}}^2$.

The present experimental constraints are the following (see 
\cite{hRll} and references therein).   
The most stringent constraint comes from the upper limit for the
flavour changing decay
$\mu\to\overline eee$:
\be
h_{e\mu}h_{ee}<3.2\times 10^{-11}\;\mrm{GeV}^{-2}\cdot M_{\Delta^{--}} ^2.
\label{hemu}
\ee
From non-observation of the decay $\mu\to e\gamma$ follows the constraint 
\be
h_{e\mu }h_{\mu\mu}  
\lsim 2 \cdot 10^{-10}\; \mrm{GeV}^{-2}\cdot M_{\Delta^{--}}^2.
\label{hemua}
\ee
From the Bhabha-scattering
cross section at SLAC and DESY the following bound  on the $h_{ee}$
coupling was established:  
\be
h_{ee}^2 \lsim 9.7 \times 10^{-6}\; \mrm{GeV}^{-2}\cdot M_{\Delta^{--}} ^2 .
\label{hee}
\ee
 For the coupling $h_{\mu\mu}$ the extra
contribution to $(g-2)_{\mu}$ yields the limit
\be
h_{\mu\mu}\lsim 2.5\cdot 10^{-5}\; \mrm{GeV}^{-2}\cdot  M_{\Delta^{--}} ^2, 
\label{hmumu}\ee 
and
the muonium transformation to
antimuonium converts into a limit 
\be
h_{ee} h_{\mu\mu } 
\lsim 5.8 \cdot 10^{-5}\; \mrm{GeV}^{-2}\cdot M_{\Delta^{--}}^2.
\label{heehmumu}
\ee
We will point out 
in this paper that in a linear collider (LC) currently under 
discussion one can
obtain much more 
stringent constraints than those quoted above by considering single
production of $\Delta^{--}$.

2.{\it Production of a single doubly charged scalar.} 
The single production of doubly
charged scalars in lepton colliders
has been previously studied by several authors
\cite{TRizzo,rizzo2,Gun}. 
The production  in $ep$
collisions at Hera was considered in \cite{acco}. 
In high energy $pp$ 
collisions at LHC the triplet $\Delta^{--}$ can be produced
via $WW$ fusion process which has been studied in \cite{meie1}.
The  rate of this process is suppressed either by the large mass of 
the right-handed gauge boson $W_R$ or by the small left-handed triplet
vacuum expectation value (vev) $v_L$.  
In $e^-\gamma$ mode the production reaction is 
\bea
e^-\g & \rightarrow & l^+ \Delta^{--}. \label{pr1} 
\eea
The photon beam can be obtained by scattering laser pulses off the 
electron beam \cite{telnov1}. 
The achievable monochromaticity and polarization rate
are comparable with the electron beam ones \cite{telnov2}.
In $e^+e^-$ collisions a single $\Delta^{--}$ can be produced in 
\bea
e^+e^- & \rightarrow & e^+ l^+ \Delta^{--}. \label{pr2}
\eea
In the $e^-e^-$ collision mode a single 
$\Delta^{--}$ will be produced in $s$-channel
anihilation. If the mass of $\Delta^{--}$ were known in advance, e.g. from 
the other LC collision modes or LHC experiments,  
one could adjust the
collision energy suitably to 
have a very large production cross section at pole. In
this way one would be 
capable to probe extremely small values of the coupling constant
$h_{ee}$, as was shown in \cite{Gun}. 

 In this paper we will 
 investigate the sensitivity of 
linear collider to the mass and couplings of a
doubly charged Higgs 
scalar in the single production  reactions (\ref{pr1}) and
(\ref{pr2}). We present our results for several collision energies
($\sqrt s= 360,\, 500,\, 800$ and 1600 GeV), 
and we take into account polarization of
the initial state particles. Unlike  previous authors 
\cite{rizzo2} we
do not use any approximations to calculate the cross sections 
of process (\ref{pr2}) and keep lepton masses explicitly in the formulae.  
We will perform our analysis in the framework of the
left-right symmetric model, 
where doubly charged scalars appear most
naturally, but our results are applicable to any model with  
scalar bileptons \cite{cuy}.

3. {\it The model.}
Let us present some basic features of the \ssu
model.
In this model leptons are assigned to 
doublets of  gauge groups $SU(2)_{L}$ and $SU(2)_{R}$ according to
their chirality:
\be
 \Psi_L = {\matr{c} \nu_e \\ e^- \ematr_L}= (2,1,-1),\;\;\;\; \Psi_R =
{\matr{c} \nu_e \\ e^-\ematr_R} = (1,2 ,-1),
\ee
 and similarly for the other families. The minimal set of fundamental
scalars consists of a bidoublet $\Phi$, 
one $SU(2)_L$ triplet $\D_L$ and one $SU(2)_R$   triplet $\D_R$. Their  quantum
numbers are $\Phi=(2,2,0),$ $\D_L=(3,1,2)$ and 
$\D_R=(1,3,2).$ This set of scalars allows for a manifest left-right symmetry
of the Lagrangian and leads to a consistent symmetry breaking scheme.

The Higgs bidoublet and triplets
\be
\begin{array}{c} {\Phi =\matr{cc}\phi_1^0&\phi_1^+\\\phi_2^-&\phi_2^0
\ematr}, \;\;\;\;
 {\Delta_{L,R} =\matr{cc}\Delta_{L,R}^+&\sqrt{2}\Delta_{L,R}^{++}\\
\sqrt{2}\Delta_{L,R}^0&-\Delta_{L,R}^+
\ematr},
\end{array}
\ee 
acquire non-vanishing vacuum expectation values given by
\be
\begin{array}{c}
{\langle\Phi\rangle=\frac{1}{\sqrt{2}}\matr{cc}\kappa_1&0\\0&\kappa_2\ematr,}
\;\;\;\;
{\langle\Delta_{L,R}\rangle
=\frac1{\sqrt{2}}\matr{cc}0&0\\v_{L,R}&0
\ematr.}
\end{array}
\ee  
The vev of the bidoublet $\Phi$
 breaks the Standard Model symmetry \sulu\  and generates Dirac mass
terms 
to fermions through the Yukawa Lagrangian  $\bar\Psi_L^i(f_{ij}\Phi +
g_{ij}\tilde\Phi)\Psi_R^j + h.c.$,  where
$\tilde\Phi=\sigma_2\Phi^*\sigma_2$. 
The left-handed triplet vev $v_L$  
is forced to be small due to its contribution to the $\rho$ parameter.
The right-triplet $\Delta_R$ breaks the $SU(2)_R\times
U(1)_{B-L}$ symmetry to
$U(1)_Y$ and at the same time the discrete $L\leftrightarrow R$
symmetry.
The vev $v_R$ gives masses to $W_R$ as well as to the doubly charged
scalars $\D^{--}_{L,R}$. 
It also provides  right-handed neutrinos with large 
Majorana mass terms via the 
 Yukawa Lagrangian
\be 
{\cal L}_Y= h_{ij}(\Psi_{iR}^TC\sigma_2\Delta_R\Psi_{jR} + 
\Psi_{iL}^TC\sigma_2\Delta_L\Psi_{jL})\;\;\;+\;{\rm h.c.},
\label{yukawa}
\ee 
where $\Psi_{iR,L}=(\nu_{iR,L},l_{iR,L})$ and $i,j$ are flavour
indices. 
This mass Lagrangian combined with the Dirac mass terms from the bidoublet
Yukawa couplings gives rise to the see-saw mechanism of neutrino masses
\cite{seesaw}, which is an attractive way to produce very light 
neutrinos.  

The processes we are interested in have their origin in the Yukawa Lagrangian
(\ref{yukawa}) 
that contains interactions between the doubly charged Higgses and
leptons. The strength of the interaction is  scaled by the unknown 
Yukawa coupling constants $h_{ij}$  which, in general,
are not flavour diagonal allowing for lepton number violating interactions. 
The present constraints on these couplings were listed above.

The two doubly charged Higgses $\D^{--}_{L,R}$  
have different chiral couplings to
leptons.   Their masses are expected to be comparable with each other 
because they derive both from similar terms of the scalar potential 
\cite{deshpande}. Since their production processes are the same 
 we shall concentrate in the following only on the production of
$\D^{--}_R.$ Therefore we assume that the electron beam is always $100\%$ 
right-handedly polarized. This is, of course, a simplification which
is motivated by the very high polarization rate achievable in the LC.
Using a right-handedly 
polarized electron beam has the benefit that at the same
time one swiches off 
most of the SM background processes which would otherwise mask
the  discovery of new particles.
  
4. {\it Production rates.}
Feynman diagrams contributing to the reaction (\ref{pr1}) are depicted
in Fig.1. 
As can be expected the dominant contribution to the cross section
comes from the diagram with a t-channel exchange of a lepton. 
It is important to
emphasize that  to get physically meaningful results one must take
into account the masses of the leptons when calculating
the helicity amplitudes of the process. The reason for that is twofold.  
Firstly, neglecting the lepton masses would make the total
cross section to diverge in the backward direction. 
Secondly, one may expect that 
with proper angular cuts one can estimate the partial cross section 
for the central region of the detector to be correct when the lepton masses 
are neglected. In this case the partial cross section is an
increasing function of the doubly charged Higgs mass $M_\D$ and
approaches a constant value when $M_{\D}$ approaches the collision
energy. This unphysical behaviour is cured when the masses
are taken into account. 
 
In Fig.2  we plot the total cross section of the process (\ref{pr1})
in the case when the final state lepton is an electron
as a function of $M_{\D}$ for two different photon beam
polarizations  and for four different collision energies 
$\sqrt{s_{e\g}}=$300, 420, 670, 1450 GeV, which correspond to 
the energies  
$\sqrt{s_{ee}}=$360, 500, 800, 1600 GeV of the $e^+e^-$ collider,
respectively. The 
Yukawa coupling constant is taken to have the value $h_{ee}=0.1.$
The cross section is large for  the whole kinematical
range of the LC. When the photon beam has linear polarization
of $\tau=-1$ the cross section is  a decreasing function of  $M_{\D}$, 
 for the  photon beam polarization
of $\tau=1$ the behaviour 
is quite the opposite, the cross section increasing with
$M_{\D}$  almost up to the kinematical threshold. 
Due to the t-channel electron exchange  the differential
cross section of the process is strongly peaked 
in the backward direction and most of the produced $\D_R^{--}$'s
have momenta almost parallel to the beam axis.

Feynman diagrams contributing to the process (\ref{pr2}) 
are depicted in Fig.3. Note that there are also crossed digrams
if both of the final state fermions are electrons.
We have neglected diagrams involving neutral Higgs bosons as
they are strongly  
suppressed due to the negligible couplings of Higgses to electrons
and   the large Higgs masses. We have also checked that
at the LC energies 
the diagrams mediated by $Z$ bosons are suppressed compared with the graphs
involving photons and can be neglected.  
Therefore, the process (\ref{pr1}) can be regarded as the subprocess of the
reaction (\ref{pr2}).

 One can estimate the total cross section
of the process (\ref{pr2}) using equivalent particle approximation (EPA)
\cite{epa} and integrating the cross section of the process (\ref{pr1})
over the photon spectrum of EPA as was done in \cite{rizzo2}. 
However, because of the delicate 
effects of small fermion  masses in this case, 
the EPA is expected to give only very
rough estimates of the total cross section and to fail to predict
further details of the process. Therefore, we have calculated the cross 
section of the process (\ref{pr2}) exactly keeping lepton masses  
non-zero both in the amplitude and in the phase space formulae.
Indeed, the behaviour of the process (\ref{pr2}) cannot be
predicted by studying the process (\ref{pr1}). The individual
contribution to the cross section of  each separate graph in Fig.3 
is huge, but they  cancel by several orders of magnitude  due to
destructive interference terms. 
The distribution of produced doubly charged particles
is peaked strongly in the forward direction.

In Fig.4  we plot the total cross section  of the process (\ref{pr2})
in the case when both final state fermions are electrons
as a function of $M_{\D}$ for collision energies of 
$\sqrt{s_{ee}}=$ 360, 500, 800, 1600 GeV assuming different longitudinal
polarizations of the  positron beam. The triplet Yukawa coupling constant
is taken to be $h_{ee}=0.1$ in these plots. 
The cross section for 
the right-handedly polarized positron beam is larger than the one
for  the left-handedly polarized 
positrons for almost all testable $M_{\D}$ range.
Close to the kinematical limit the difference can be more than 
one  order
of magnitude. 
Only for relatively small $M_\D$ values the situation is 
the opposite but 
in this case the cross sections are comparable in size.
Clearly, with the chosen parameters one can test $\D_R^{--}$ 
masses almost up to the kinematical threshold.

In order to discover $\D_R^{--}$ at the LC one has to detect its decay
products. The decays of doubly charged Higgs bosons have been studied e.g.
in \cite{meie1}. With reasonable assumptions on the
masses of Higgs bosons 
and various mixing angles the main decay modes were found to be
$\D_R^{--}\rightarrow l_1^-l^-_2$ and $\D_R^{--}\rightarrow W^-_R W^-_R,$
where $l_{1,2}$ denote leptons. 
Since for the $\Delta_R$ mass range we are considering the latter decay is
kinematically forbidden, 
we assume in the following that $\D_R^{--}$ decays $100\%$
to leptons. The experimental signature of the decay
is then the same sign lepton pair with no missing energy, including 
lepton number violating final states, e.g. $\mu^-\mu^-$  or $e^-\mu^-$.

As can be seen in Fig.5.,
 for a light  $\D_R^{--}$ the distribution of a produced lepton $l^-$ is
peaked in  the backward (process (\ref{pr1})) or forward (process (\ref{pr2}))
direction, for larger masses the distribution becomes more flat.
This is an expected behaviour since the heavier the particles, the
 less boosted they are
along the beam axis. Assuming the coverage of the detector to be 
$|\cos\theta|<0.95$ one can detect about $60\%$ of the 
final state leptons if 
$M_{\D}=100$ GeV and considerably more for larger masses.

 The opening angle between the two produced leptons strongly depends
on the mass of
$\D_R^{--}.$ For heavy $\D_R^{--}$ 
the final state leptons are almost back to back
while for 100 GeV  Higgs boson the opening angle is restricted to be 
$\cos\theta_{12}\lsim 0.5.$ This implies that one of the leptons 
can  always be detected.  
 Taking into account the rather small losses of the 
decay products,
clear experimental signatures (lepton number violation, no missing energy)
 and almost no background from the SM, the doubly charged Higgs 
is hard to  miss  in  LC if it is produced. 

5. {\it Sensitivity to $|\Delta L|=2$ couplings.}
 Let us finally analyse the
sensitivity of the LC to the triplet Yukawa couplings $h_{ij}.$
The couplings can be tested either in the production 
processes or in the decays of the Higgs bosons.
Since our first  interest is  the production of $\D_R^{--}$ 
we will assume that 
the Higgs boson mass is in the LC energy range. If $\D_R^{--}$ will not be
seen in the collider experiments one will be able to put an upper bound on the
couplings.   

The present constraints (\ref{hee}), (\ref{hmumu}) and (\ref{heehmumu}) do not
yield any essential restriction 
on the diagonal couplings $h_{ee}$ and $h_{\mu\mu}$
for the scalar 
masses above the electroweak scale, and as large values of the couplings
as ${\cal O}(1)$ are allowed. The off-diagonal coupling $h_{e\mu}$ is instead
restricted by (\ref{hemu}) 
well below unity in the whole mass range covered by the
LC. At the LC one will be 
able via the reactions (\ref{pr2}) and (\ref{pr1}) to
improve these constraints substantially. Furthermore, in addition to  the
non-diagonal
$h_{e\mu}$ coupling one will be able to 
probe also the non-diagonal coupling $h_{e\tau}$,
for which  there are no bounds  at all from low energy processes. 

Let us first consider the reaction (\ref{pr1}).  
The primary lepton created in the process will remain undetected as it is
radiated almost parallel to  the beam axis. 
One cannot tell 
whether this particle is a positron, antimuon or antitau.  Therefore,
the quantity which one can test in the reaction is actually the sum
$ h_{ee}^2+h_{e\mu}^2+h_{e\tau}^2$. 
The upper bound obtained for this sum is, of
course, the upper bound  of  each individual term of the sum separately.

Assuming the integrated luminosities of $e^-\g$ collisions to be
$L=5,$ 10, 20, 40 fb$^{-1}$ 
and that for the discovery of $\D_R^{--}$ one needs
ten events, we obtain   the upper bounds plotted in Fig.6. 
As one can see from the
figure, the sensitivity of the LC on the quantity
$(h_{ee}^2+h_{e\mu}^2+h_{e\tau}^2)^{1/2}$ 
is on the level of $10^{-3}$ almost up
to the threshold value of the
$\D_R^{--}$ mass. In other words,
\be
h_{ee},\, h_{e\mu},\, h_{e\tau}\lsim 10^{-3}
\ee
for $M_{\Delta^{--}}\lsim \sqrt{s_{e\gamma}}$. 
Among the present constraints only
(\ref{hemu}), also presented in the plot, 
competes with these bounds and does so only at
the low mass values. For the coupling $h_{e\tau}$ no bounds exist from the
present experiments.

  For the same $\D_R^{--}$ mass, the cross section of the process
(\ref{pr2}) is  roughly two orders of magnitude 
smaller than the cross section of 
the process (\ref{pr1}), implying that the constraints obtained for
$(h_{ee}^2+h_{e\mu}^2+h_{e\tau}^2)^{1/2}$ are correspondingly weaker,
although  the higher luminosities  $L=$ 20, 50, 100, 200 fb$^{-1}$
of $e^+e^-$ slightly  compensate the lack in cross section.
The resulting bounds are presented in Fig.7.

6.{\it Summary.}
We have studied in the framework of the left-right symmetric model the single 
 production of  a doubly
charged Higgs boson via the reactions 
$e^-e^+  \rightarrow  e^+ l^+
\Delta^{--}$ and
$
e^-\g  \rightarrow  l^+ \Delta^{--}$ at the LC. 
We found that the testable range of
$\D_R^{--}$ mass extends  almost up to the collision energy. 
A negative search of
the double charged scalar 
will lead to constraints for the lepton number violating
Yukawa couplings substantially more stringent than the present ones.

\subsection*{Acknowledgement}

We thank F. Botella and A. Santamaria for clarifying discussions.
G.B. acknowledges the Spanish Ministry of
Foreign Affairs for a MUTIS fellowship and M.R. thanks the
Spanish Ministry of Science and Education for a postdoctoral
grant at the University of Valencia. This work is supported by CICYT under 
grant AEN-96-1718 and by the Academy of Finland.

\newpage

{\large\bf Figure caption}

\vspace{0.5cm}

\begin{itemize}

\item[{\bf Fig.1}] 
Feynman diagrams of the process $e^-\g\rightarrow l^+\D^{--}$
 in the left-right model.
\\
\item[{\bf Fig.2}]
Total cross section of the process $e^-\g\rightarrow e^+\D^{--}$
as a function of Higgs boson mass $M_{\D}$ for different beam polarizations
and collision energies as indicated in figure.
\\
\item[{\bf Fig.3}]
Feynman diagrams contributing to the process 
$e^-e^+\rightarrow e^+l^+\D^{--}$ in the left-right model.
\\
\item[{\bf Fig.4}] 
Total cross section of the process $e^-e^+\rightarrow e^+e^+\D^{--}$ 
as a function of Higgs boson mass $M_{\D}$ for different beam polarizations
and collision energies as indicated in figure.
\\
\item[{\bf Fig.5}] 
Angular distribution of final state leptons produced in
decays of doubly charged Higgs boson $\D_R^{--}$. Production processes and 
masses of the Higgs  boson are shown in figure. 
\\
\item[{\bf Fig.6}] 
Achievable constraints on triplet Yukawa couplings from the process
$e^-\gamma\rightarrow l^+\Delta_R^{--}$ for different collision energies 
as functions of the scalar mass.
The most stringent 
present constraint from low energy experiments is also drawn.
\\
\item[{\bf Fig.7}] 
Achievable constraints on triplet Yukawa couplings from the process
$e^-e^+\rightarrow e^+l^+\Delta_R^{--}$ for different collision energies
as functions of the scalar mass.
The most stringent 
present constraint from low energy experiments is also drawn.

\end{itemize}

\newpage

\begin{figure*}
\begin{center}
 \mbox{\epsfxsize=16cm\epsfysize=10cm\epsffile{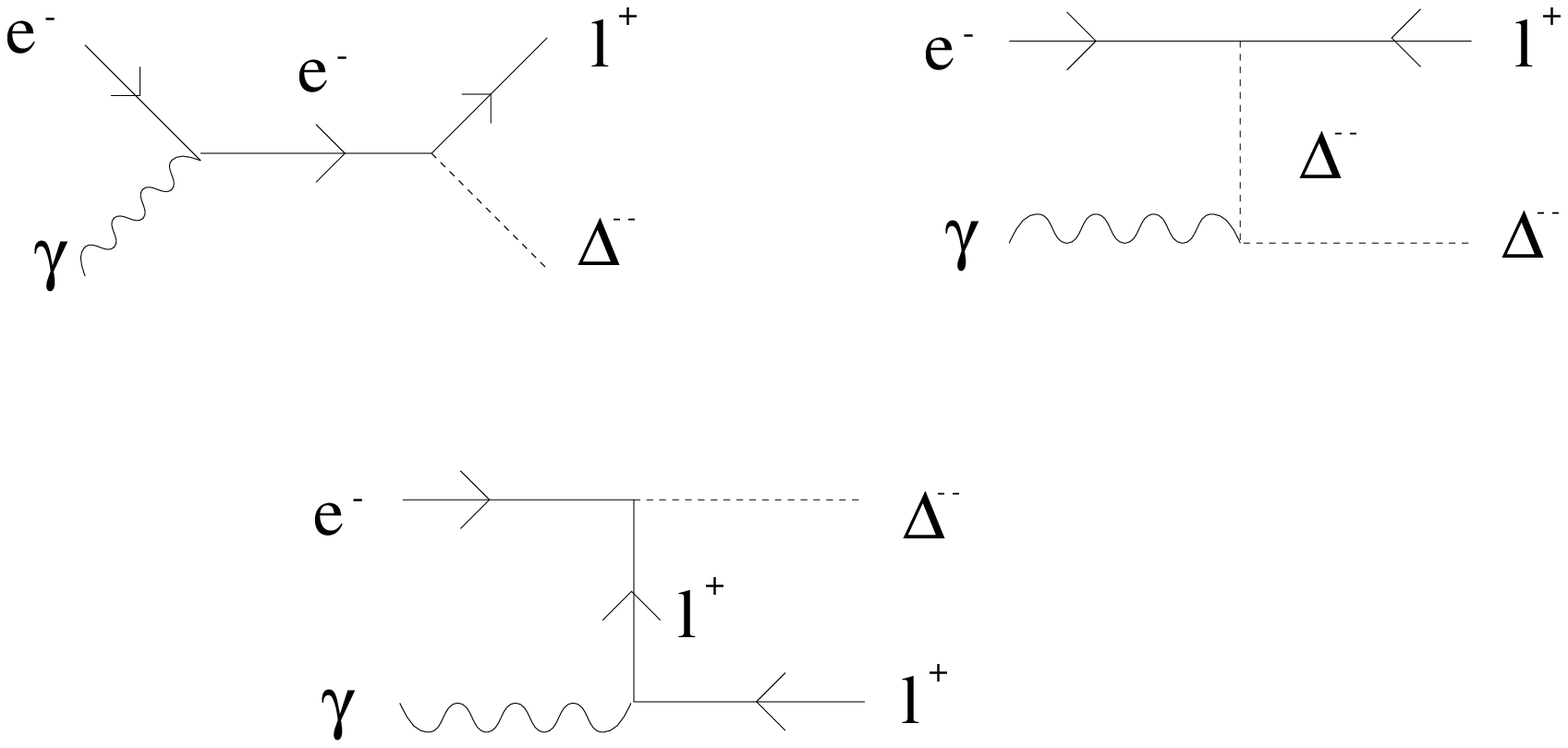}}
\caption{}
\end{center}
\end{figure*}

\begin{figure*}
\begin{center}
 \mbox{\epsfxsize=16cm\epsfysize=16cm\epsffile{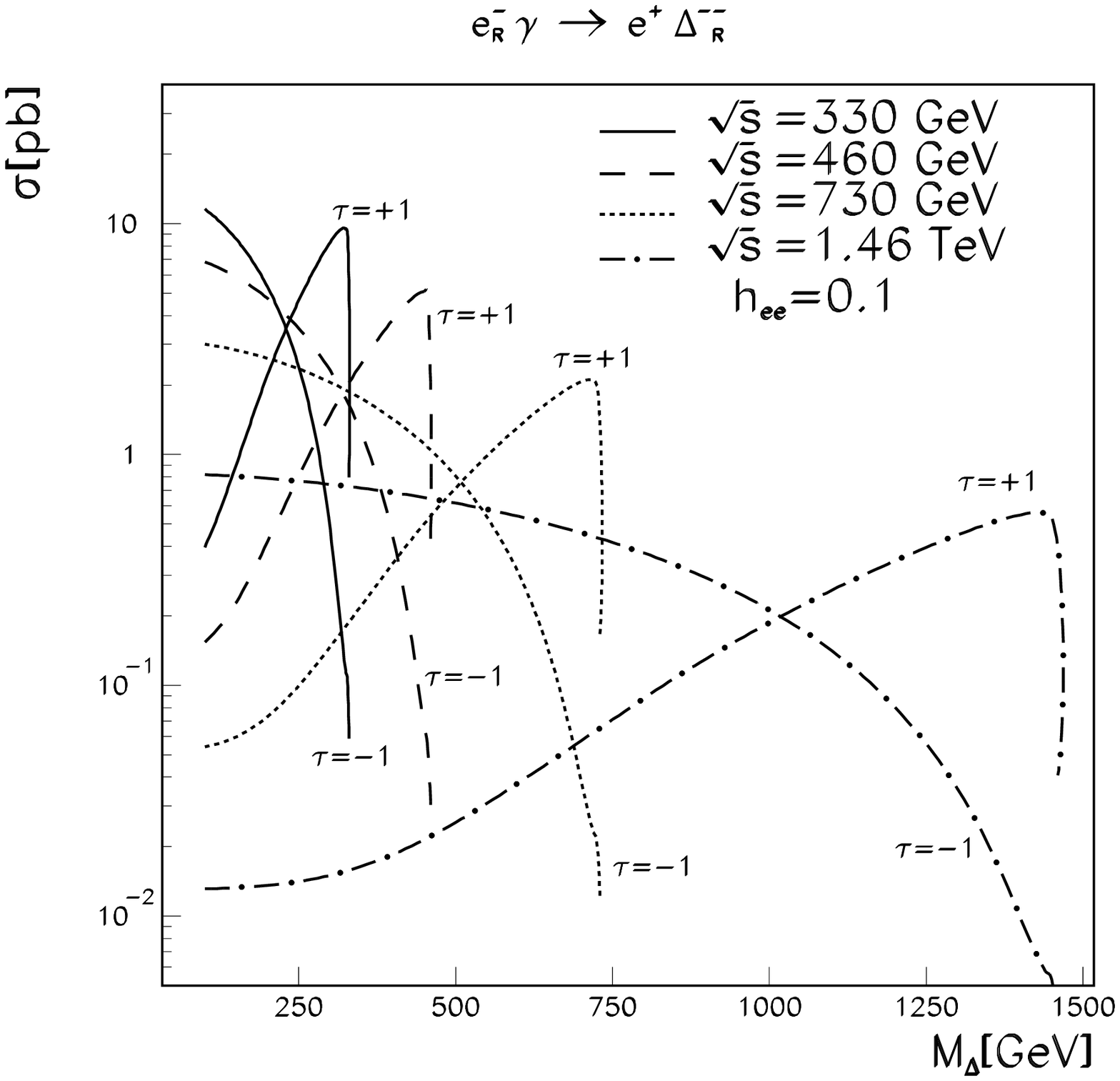}}
\caption{}
\end{center}
\end{figure*}

\begin{figure*}
\begin{center}
 \mbox{\epsfxsize=16cm\epsfysize=10cm\epsffile{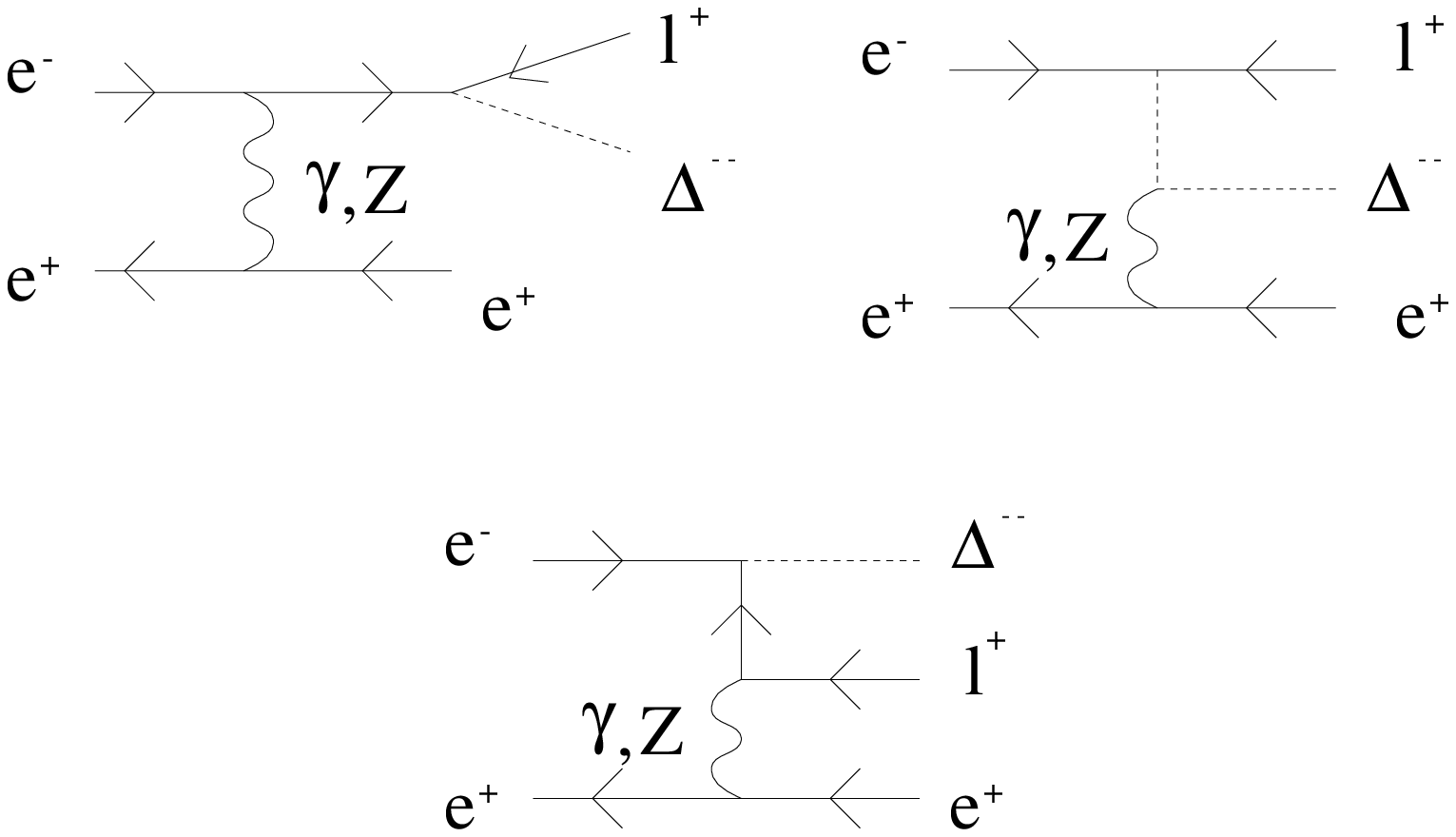}}
\caption{}
\end{center}
\end{figure*}

\begin{figure*}
\begin{center}
 \mbox{\epsfxsize=16cm\epsfysize=16cm\epsffile{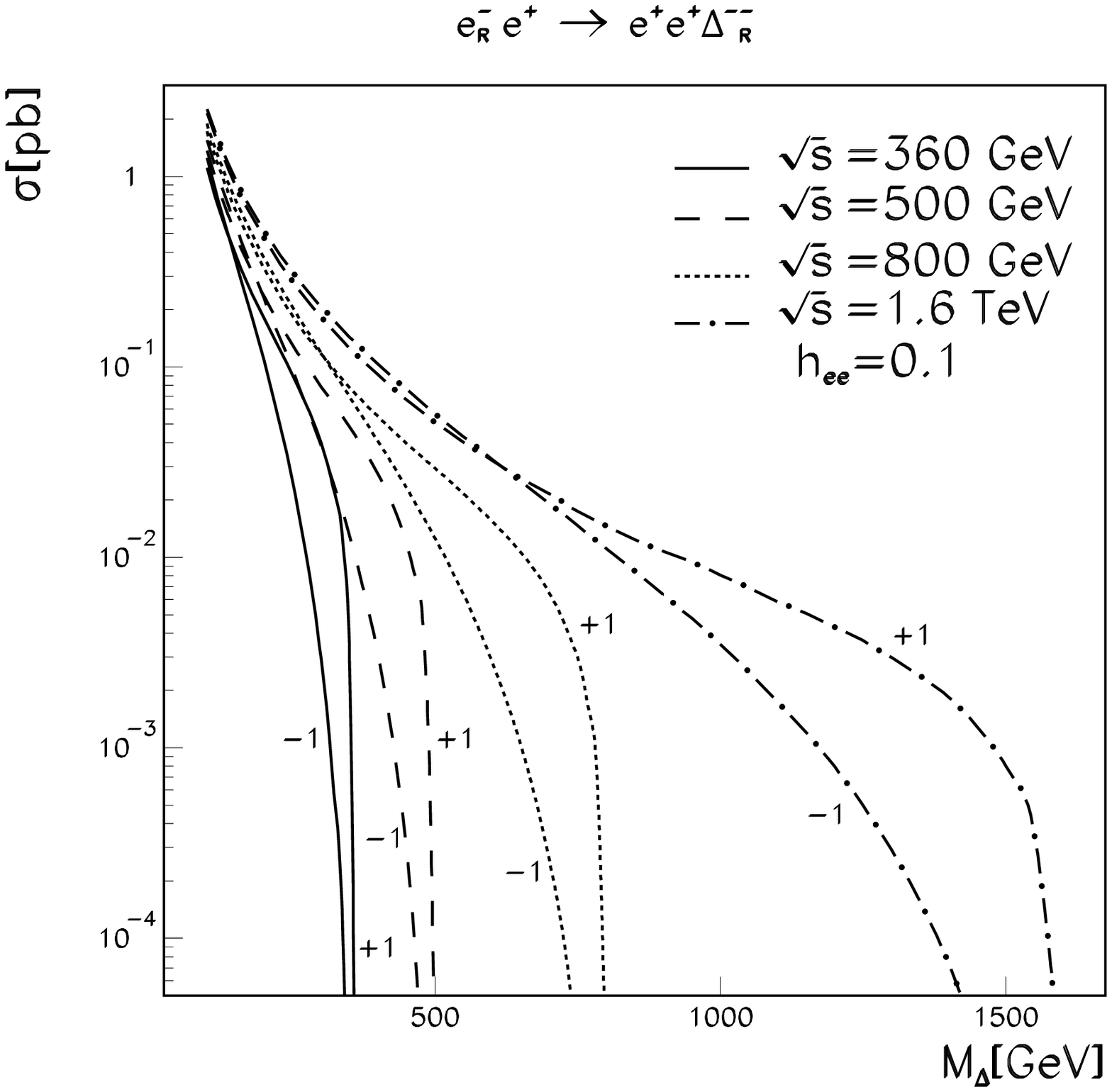}}
\caption{}
\end{center}
\end{figure*}

\begin{figure*}
\begin{center}
 \mbox{\epsfxsize=16cm\epsfysize=16cm\epsffile{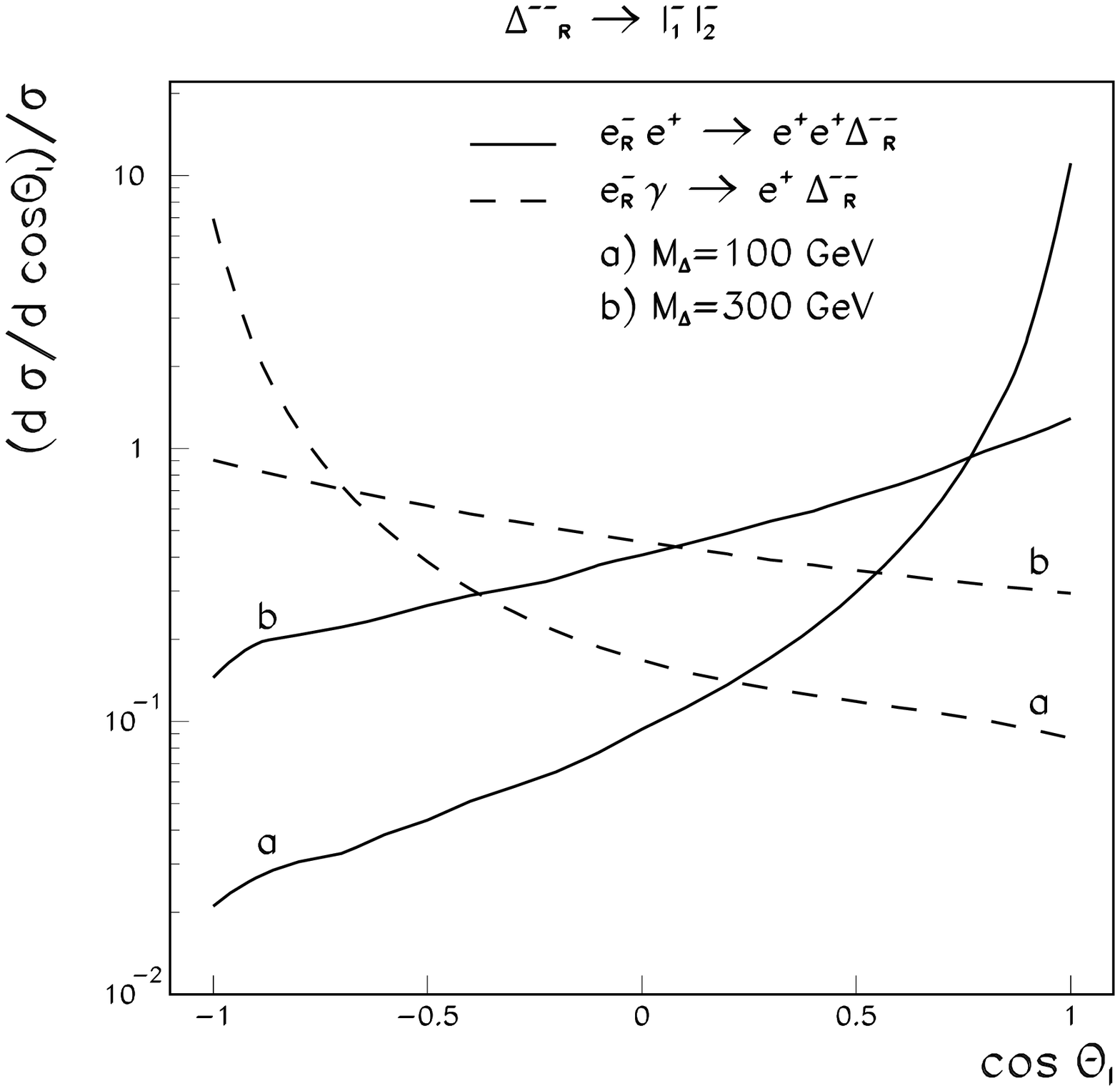}}
\caption{}
\end{center}
\end{figure*}

\begin{figure*}
\begin{center}
 \mbox{\epsfxsize=16cm\epsfysize=16cm\epsffile{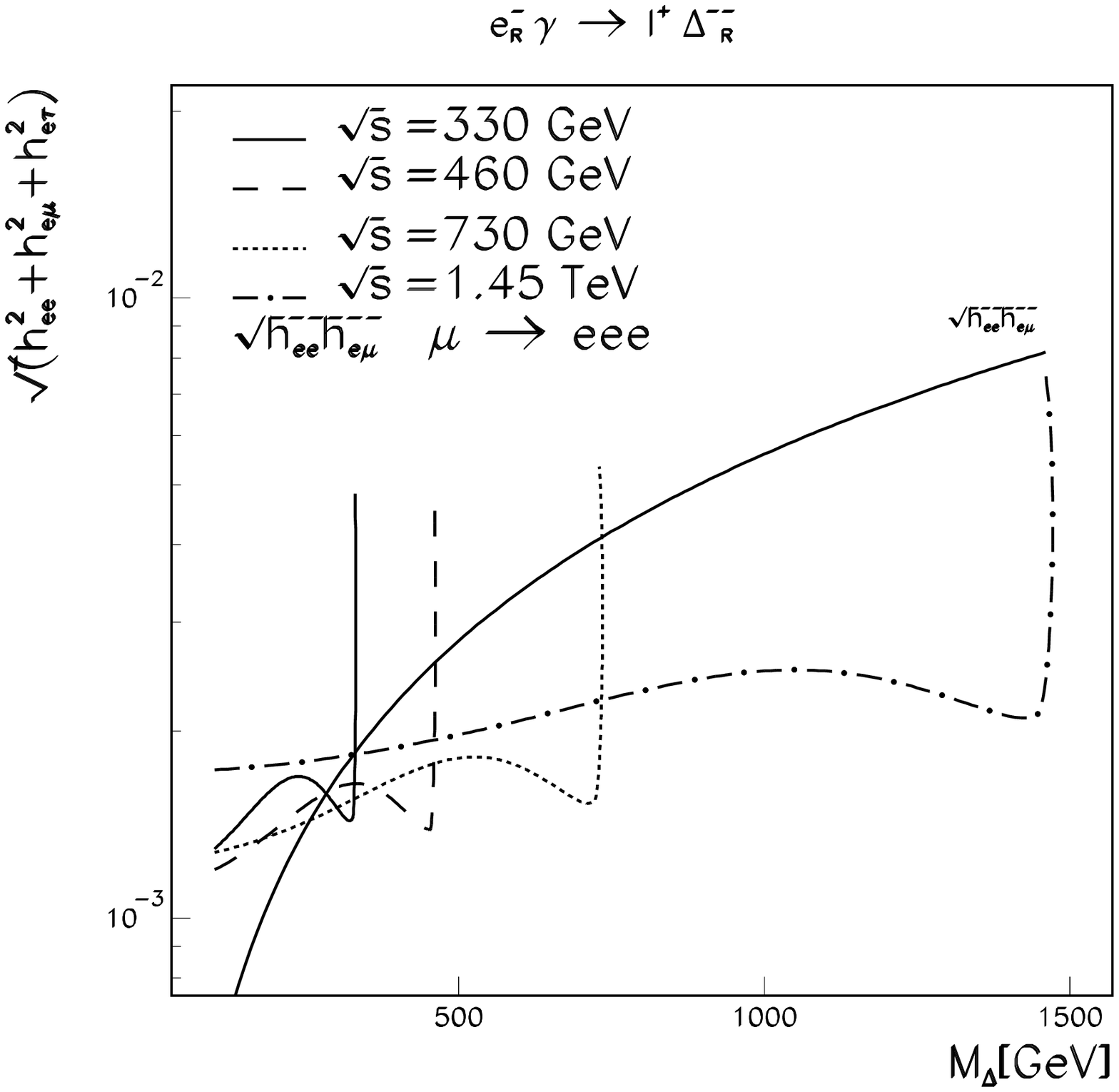}}
\caption{}
\end{center}
\end{figure*}

\begin{figure*}
\begin{center}
 \mbox{\epsfxsize=16cm\epsfysize=16cm\epsffile{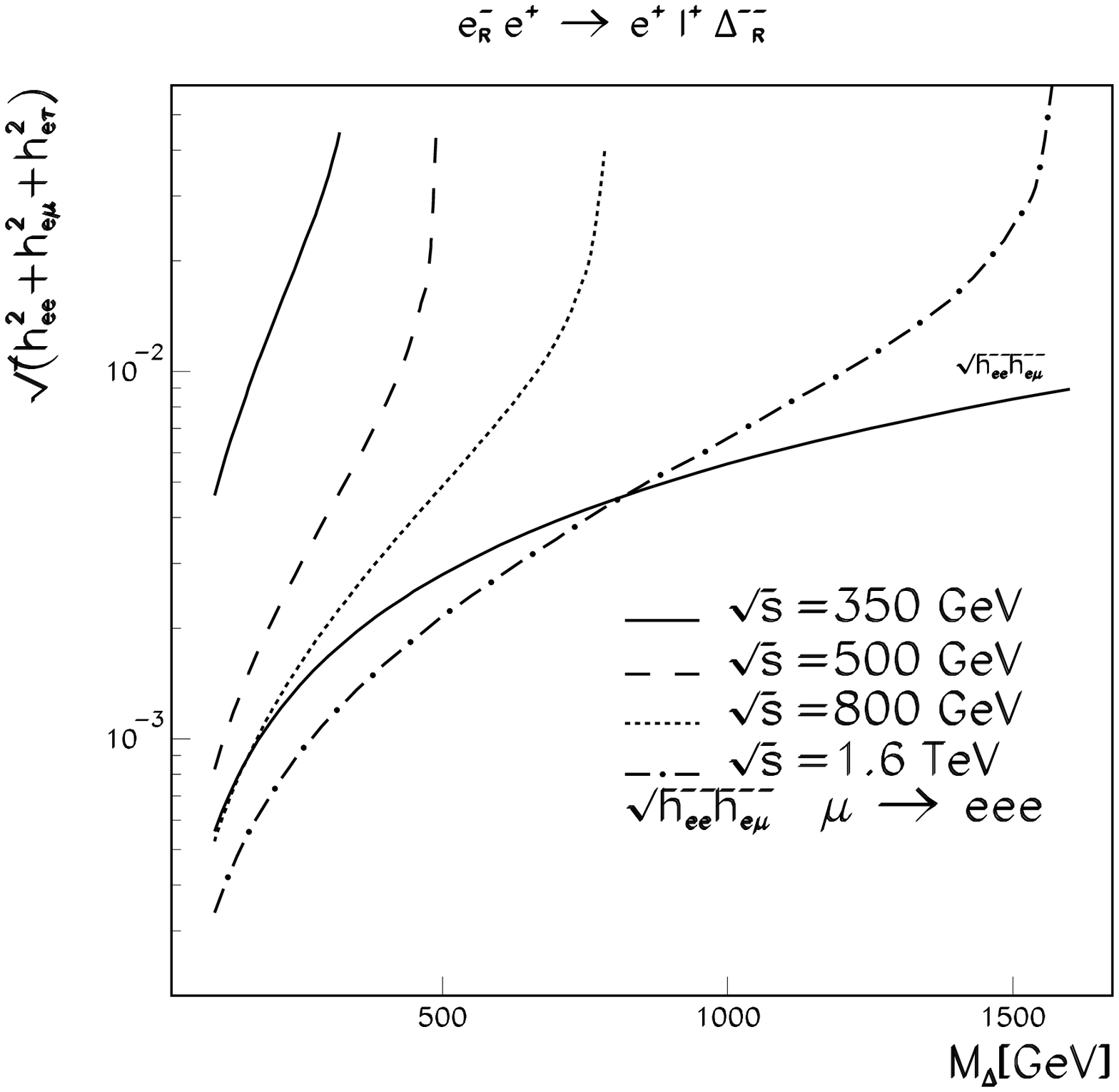}}
\caption{}
\end{center}
\end{figure*}

\end{document}